\documentclass[twocolumn,pra,superscriptaddress]{revtex4}
\pdfoutput=1
\usepackage{setspace}
\usepackage{amsfonts,amsmath,amssymb}
\usepackage{txfonts}
\usepackage{graphicx, color}
\usepackage{verbatim}
\usepackage{hyperref}
\usepackage{bm}

\begin{document}

\title{C-Field Methods for Non-Equilibrium Bose Gases}

\author{Matthew J. Davis}
\affiliation{The University of Queensland, School of Mathematics and Physics, Brisbane, Queensland 4072, Australia} 
\author{Tod M. Wright} 
\affiliation{The University of Queensland, School of Mathematics and Physics, Brisbane, Queensland 4072, Australia} 
\author{P. Blair Blakie}
\affiliation{Jack Dodd Centre for Quantum Technology, Department of Physics, University of Otago, Dunedin, New Zealand}
\author{Ashton S. Bradley}
\affiliation{Jack Dodd Centre for Quantum Technology, Department of Physics, University of Otago, Dunedin, New Zealand}
\author{Rob J. Ballagh}
\affiliation{Jack Dodd Centre for Quantum Technology, Department of Physics, University of Otago, Dunedin, New Zealand}
\author{Crispin W. Gardiner}
\affiliation{Jack Dodd Centre for Quantum Technology, Department of Physics, University of Otago, Dunedin, New Zealand}

\begin{abstract}
We review c-field methods for simulating the non-equilibrium dynamics of degenerate Bose gases beyond the mean-field Gross--Pitaevskii approximation.  We describe three separate approaches that utilise similar numerical methods, but have distinct regimes of validity.  Systems at finite temperature can be treated with either the closed-system projected Gross--Pitaevskii equation (PGPE), or the open-system stochastic projected Gross--Pitaevskii equation (SPGPE).  These are both applicable in quantum degenerate regimes in which thermal fluctuations are significant.  At low or zero temperature, the truncated Wigner projected Gross--Pitaevskii equation (TWPGPE) allows for the simulation of systems in which spontaneous collision processes seeded by quantum fluctuations are important.  We describe the regimes of validity of each of these methods, and discuss their relationships to one another, and to other simulation techniques for the dynamics of Bose gases.  The utility of the SPGPE formalism in modelling non-equilibrium Bose gases is illustrated by its application to the dynamics of spontaneous vortex formation in the growth of a Bose--Einstein condensate. 
\end{abstract}

\maketitle

\section{Introduction}
It may seem counter-intuitive that the dynamics of Bose--Einstein condensation, a phenomenon that occurs due to \emph{quantum} statistics,  can often be well-described by an equation of motion for a \emph{classical} field.  However, there are in fact  two distinct classical regimes~\cite{polkovnikov_10} that can potentially be confused.  The first is the \emph{classical-particle} regime ---  where the de Broglie wavelength of the atoms in a gas is much smaller than the typical interparticle spacing, and the particles can be treated as `billiard balls'.  The second is the \emph{classical-field} regime, in which the modes of a bosonic quantum field are highly occupied ($\langle \hat{N}_k \rangle \gg 1$), such that the discrete set of integer mode occupations can be approximated by a continuum.   This occurs near quantum degeneracy for the ultra-cold Bose gas,  and in these circumstances a description using  a classical wave equation may be  appropriate.  There are two cases worth distinguishing: The Gross--Pitaevskii equation (GPE)~\cite{gross_61,pitaevskii_61} for a pure Bose--Einstein condensate (BEC) at zero temperature approximates the highly-occupied condensate as a classical field, analogously to the representation of the quantum field of laser light by a classical electromagnetic wave~\cite{cohen-tannoudji_book_92}.  However, just as the long-wavelength limit of the Planck distribution of blackbody radiation --- in which the distinction between individual field quanta is unnecessary --- can be described by the classical Rayleigh--Jeans law~\cite{pathria_book_96}, the GPE can also provide, within appropriate limits, a description of the finite-temperature atomic Bose field.  This Chapter describes methodologies in which an equation of motion for a classical field provides a beyond-mean-field description of the degenerate Bose gas.

A number of authors have proposed that a classical-field treatment of the Bose gas is appropriate to describe the dynamics of condensate formation.  Near degeneracy, a quantum Boltzmann equation description becomes inadequate due to the growing coherences between the modes.   Kagan, Svistunov, and Shlyapnikov suggested that when the inequality $N_k \equiv \langle \hat{a}_k^{\dag}  \hat{a}_k \rangle \gg 1$ is satisfied for a large number of the low-energy modes of the gas, the Bose field is more appropriately described by a Gross--Pitaevskii equation~\cite{svistunov_91,kagan_svistunov_92a,kagan_svistunov_92b,kagan_svistunov_94,kagan_svistunov_97}.  Other authors, including Stoof~\cite{stoof_99} and Gardiner \emph{et al.}~\cite{gardiner_anglin_02,gardiner_davis_03}, came to realise the need for a classical-field description from different starting points.

The first numerical simulations of the GPE demonstrating the kinetics of Bose condensation in a classical-field model were performed by Damle \emph{et al.}~\cite{damle_majumda_96}.  This was followed by work by Marshall  \emph{et al.}~\cite{marshall_new_99}, Stoof \emph{et al.}~\cite{stoof_bijlsma_01, duine_stoof_01}, G\'{o}ral  \emph{et al.}~\cite{goral_gajda_01a}, Sinatra  \emph{et al.}~\cite{sinatra_lobo_01},  Davis  \emph{et al.}~\cite{davis_morgan_01,davis_morgan_02,blakie_davis_05a}, and Berloff and Svistunov~\cite{berloff_svistunov_02}, who all simulated the Bose gas at finite temperature using methods based on the GPE.  Collectively, this type of approach has become known as the `classical-field method'~\cite{blakie_bradley_08,brewczyk_gajda_07,stoof_99}.  While the simplest variants of this method entirely neglect the effects of quantum fluctuations, they have the significant advantage that they treat the classical fluctuations of the field non-perturbatively, and hence can be applied in the critical regime around the Bose-condensation phase transition~\cite{arnold_moore_01,kashurnikov_prokofev_01,andersen_04,davis_morgan_03,davis_blakie_06,bezett_blakie_09a} and the fluctuation regime of the two-dimensional Bose gas~\cite{prokofev_ruebenacker_01,prokofev_svistunov_02,bisset_davis_09,foster_blakie_10}.  The related truncated Wigner methods \cite{steel_olsen_98,sinatra_lobo_02,polkovnikov_03a,isella_ruostekoski_06,blakie_bradley_08} incorporate the leading-order effects of quantum fluctuations into a GPE-like description~\cite{polkovnikov_03}.

In this Chapter we briefly outline the derivation of  the stochastic projected Gross--Pitaevskii equation (SPGPE) formalism --- for a recent, more extensive review see Ref.~\cite{blakie_bradley_08}.  The SPGPE is an equation of motion for a classical field (\mbox{c-field}) describing the low-energy, highly-occupied modes of a Bose gas, coupled to a bath of high-energy atoms that is assumed to be close to thermal equilibrium.  It offers a powerful framework for the study of both equilibrium correlations --- including condensation, anomalous correlations, and critical fluctuations --- and non-equilibrium dynamics of the finite-temperature Bose gas.  We also discuss two other related methods:  The projected Gross--Pitaevskii equation (PGPE) formalism is obtained from the SPGPE upon neglecting the coupling of the c-field to the bath, and the projected truncated-Wigner method (TWPGPE) incorporates the leading-order effects of quantum fluctuations~\cite{polkovnikov_03} at low temperatures, where such fluctuations may be important.  We describe each of these methods and their regimes of validity, and discuss their relationship to the work of other authors.  Finally we describe the application of the SPGPE to the modelling of spontaneous vortex formation in condensate growth.

\section{Methodology}
\subsection{Outline of Derivation}
The classical Rayleigh--Jeans law for thermal black-body radiation is a good approximation for the long-wavelength modes of the electromagnetic field, but results in the well-known ultra-violet catastrophe at short wavelengths~\cite{pathria_book_96}.  This suggests that a classical treatment may provide a good approximate description of the atomic Bose field when restricted to the low-energy field modes.  We therefore divide the Bose field operator $\hat{\Psi}(\mathbf{r})$ into a low-energy part that will be treated classically (the coherent or c-field region $\mathbf{C}$), and a high-energy part that will be treated quantum mechanically  (the incoherent region $\mathbf{I}$)
\begin{equation}
\hat{\Psi}(\mathbf{r}) = \hat{\Psi}_{\mathbf{C}}(\mathbf{r})  + \hat{\Psi}_{\mathbf{I}}(\mathbf{r}).
\end{equation}
This division is effected by defining a projection operator onto the c-field  region
\begin{equation}
 \mathcal{P}_{\mathbf{C}}\{f(\mathbf{r})\} = \sum_{n\in\mathbf{C}}\; \varphi_n(\mathbf{r}) \int \! d\mathbf{r}'\, \varphi^*_n(\mathbf{r}') f(\mathbf{r}'),
\end{equation}
such that
\begin{align}
\hat{\Psi}_{\mathbf{C}}(\mathbf{r})  &= \mathcal{P}_{\mathbf{C}} \{ \hat{\Psi}(\mathbf{r}) \} = \sum_{n\in\mathbf{C}} \hat{a}_n \varphi_n(\mathbf{r}),
\end{align}
where the $\varphi_n(\mathbf{r})$ are the eigenvectors of the appropriate single-particle Hamiltonian, with eigenvalues $\varepsilon_n$.   The division between the $\mathbf{C}$ and $\mathbf{I}$ regions is made at a cutoff energy $E_{\rm cut}$, which is chosen such that the modes in $\mathbf{C}$ are classical, that is $\langle \hat{N}_k \rangle \gg 1$ for the highest energy mode in $\mathbf{C}$.  The particular choice of $E_{\rm cut}$ will be dependent on the thermodynamic parameters of the initial state of the system, but we require that the physical observables we calculate should be insensitive to the exact value chosen for the cutoff~\cite{davis_blakie_06}.

Our  goal is to derive a computationally tractable equation of motion for the c-field region.  To achieve this we make use of the methods of open quantum systems~\cite{gardiner_zoller_book_10}, and treat the incoherent region $\mathbf{I}$ as a thermal and diffusive reservoir (or \emph{bath}) to which the field $\hat{\Psi}_{\mathbf{C}}(\mathbf{r})$ is coupled~\cite{gardiner_davis_03}.  We assume that the bath density operator is thermal and quantum Gaussian~\cite{gardiner_zoller_book_10}, so that all operator products factorise into products of one-body correlation functions.  Furthermore, we assume that these correlations are well represented by a semiclassical description in terms of a one-body Wigner function $F_\mathbf{I}(\mathbf{r},\mathbf{K})$~\cite{castin_chapter_01}.   In principle, an equation of motion for $F_\mathbf{I}(\mathbf{r},\mathbf{K})$ could be derived using the methods of kinetic theory, and this is a topic of current research.  Here we assume thermal equilibrium such that 
\begin{equation}
F_\mathbf{I}(\mathbf{r},\mathbf{K}) = 
\begin{cases}
1/\{\exp[(\tilde{\varepsilon}(\mathbf{r},\mathbf{K}) - \mu)/k_\mathrm{B} T] -1\}
&
\tilde{\varepsilon}(\mathbf{r},\mathbf{K}) > E_{\rm cut},\\
0
&
\tilde{\varepsilon}(\mathbf{r},\mathbf{K}) \le E_{\rm cut},
\end{cases}
\end{equation}
where $\tilde{\varepsilon}(\mathbf{r},\mathbf{K})$ is the semiclassical energy~\cite{dalfovo_giorgini_99} of a particle with position $\mathbf{r}$ and wavevector $\mathbf{K}$.  The $\mathbf{I}$ region is then completely characterised by its chemical potential $\mu$ and temperature $T$ (and any other appropriate thermodynamic Lagrange multipliers~\cite{bradley_gardiner_08}), and these parameters will arise in the equation of motion for the classical field. 

A master equation for the coherent-region density operator can be obtained by tracing out the incoherent region, using standard methods from quantum optics~\cite{gardiner_zoller_book_10}.  A `high-temperature' approximation is then made for the master equation, requiring that all the eigenfrequencies of the $\mathbf{C}$-region evolution are small compared to the temperature; i.e., $\hbar|\epsilon_j|/k_\mathrm{B} T \ll 1$~\cite{gardiner_davis_03} (in practice we find $\hbar|\epsilon_j| \lesssim \mu+E_{\rm cut}$~\cite{wright_ballagh_08}).  
The master equation can then be mapped to a generalised Fokker--Planck equation for the Wigner quasi-probability distribution using the operator correspondences described in Refs.~\cite{gardiner_zoller_book_10,blakie_bradley_08}. The binary interaction term in the standard Bose-field Hamiltonian leads to terms involving third-order derivatives of the Wigner distribution with respect to the phase-space variables~\cite{steel_olsen_98}, so there is no exact representation of the resulting Fokker--Planck equation in terms of a stochastic differential equation (SDE).  However, for highly occupied c-field modes we may make the truncated Wigner approximation (TWA), in which the third-order derivatives are neglected~\cite{steel_olsen_98}.  The resulting (Stratonovich form~\cite{gardiner_book_04}) stochastic differential equation for the classical-field $\Phi(\mathbf{r})$ is the SPGPE
\begin{subequations}
\begin{align}
\!\!\!d\Phi(\mathbf{r})
=& \,\mathcal{P}_{\mathbf{C}}\left\{ 
-\frac{i}{\hbar} H_0 \Phi(\mathbf{r})\right. dt 
\label{eq:SGPE_line1}\\
&\!\!+
\frac{G(\mathbf{r})}{k_\mathrm{B} T} (\mu - H_0)\Phi(\mathbf{r})\, dt 
+ dW_G(\mathbf{r},t)
\label{eq:SGPE_line2}\\
&\!\!+
\left.
\int\!\!{d\mathbf{r}^\prime}M(\mathbf{r}-\mathbf{r}^\prime)\frac{ i\hbar\nabla\cdot \mathbf{j}_\mathbf{C}(\mathbf{r}^\prime)}{k_\mathrm{B}T} \Phi(\mathbf{r})dt
+i\Phi(\mathbf{r})dW_{M}(\mathbf{r},t)
\right\}\!,
\label{eq:SGPE_line3}
\end{align}
\label{eq:SPGPE_all}
\end{subequations}
where 
\begin{equation}
H_0 = -\frac{\hbar^2}{2m} \nabla^2 + V_{\mathrm{ext}}(\mathbf{r})
+ g |\Phi(\mathbf{r})|^2,
\end{equation}
is the Gross--Pitaevskii operator, with $g$ the three-dimensional interaction strength. All terms on the RHS of the SPGPE involve the c-field projection operator $\mathcal{P}_{\mathbf{C}}$, which formally restricts the dynamics of  $\Phi(\mathbf{r})$ to the region $\mathbf{C}$.  Aside from the projection, the first line of the SPGPE, Eq.~(\ref{eq:SGPE_line1}), is the standard zero-temperature GPE. 

Equations (\ref{eq:SPGPE_all}b)--(\ref{eq:SPGPE_all}c) describe the coupling of the c-field to the $\mathbf{I}$ region. Equation (\ref{eq:SGPE_line2}) corresponds to growth processes, in which two $\mathbf{I}$-region atoms collide and one is scattered into $\mathbf{C}$, and the corresponding time-reversed processes.  The local c-field amplitude grows if the  value of $H_0 \Phi(\mathbf{r})$ is smaller than  $\mu  \Phi(\mathbf{r})$, and vice versa.
 The \emph{growth rate} $G(\mathbf{r})$ is given by a collision integral over the $\mathbf{I}$ region --- see Eq.~(171) of Ref.~\cite{blakie_bradley_08}. 

Equation~(\ref{eq:SGPE_line3}) corresponds to scattering processes in which a particle in the $\mathbf{I}$ region scatters off one in the $\mathbf{C}$ region, leaving the total number in each region unchanged.  These processes couple the bath to the divergence of the c-field current 
\begin{equation}
\mathbf{j}_{\mathbf{C}}(\mathbf{r})\equiv \frac{i\hbar}{2m}\Big([\nabla \Phi^*(\mathbf{r})] \Phi(\mathbf{r})
- \Phi^*(\mathbf{r})\nabla \Phi(\mathbf{r})\Big),
\end{equation}
and are characterised by the scattering amplitude $M(\mathbf{r}-\mathbf{r}^\prime)$, which is given by another collision integral [Eq.~(174) of Ref.~\cite{blakie_bradley_08}].

The SPGPE involves two explicit stochastic terms: a complex noise process $dW_G(\mathbf{r},t)$ associated with growth, and a real noise process $dW_{M}(\mathbf{r},t)$ associated with scattering.  The non-zero correlations of these Wiener increments are
\begin{align}
\langle dW_G^*(\mathbf{r},t)dW_G(\mathbf{r}',t) \rangle =& 2 G(\mathbf{r}) \delta_\mathbf{C}(\mathbf{r},\mathbf{r}') dt,\\
\langle dW_{M}(\mathbf{r},t)dW_{ M}(\mathbf{r}^\prime,t)\rangle =&2M(\mathbf{r}-\mathbf{r}^\prime) dt,
\end{align}
where $ \delta_\mathbf{C}(\mathbf{r},\mathbf{r}^\prime) \equiv \sum_{n\in\mathbf{C}}  \varphi_n(\mathbf{r})\varphi^*_n(\mathbf{r}^\prime)$ is the kernel of the projection operator $\mathcal{P}_\mathbf{C}$ and acts as a Dirac delta function within the c-field region.

As the SPGPE is derived within the TWA, the initial condition $\Phi(\mathbf{r},t=0)$ should in principle include a representation of quantum fluctuations; i.e., half a particle of noise per mode~\cite{blakie_bradley_08}.  Distinct samples of this initial `vacuum noise' and the dynamical noise processes generate distinct SPGPE trajectories, and in the Wigner formalism moments of the c-field $\Phi(\mathbf{r})$ calculated from an ensemble of such trajectories correspond to quantum-statistical expectation values of symmetrically ordered field-operator products~\cite{steel_olsen_98}.  In practice, the evolution of the field is rapidly dominated by classical fluctuations which overwhelm the contributions of the vacuum noise, and the formal operator-ordering corrections of the Wigner interpretation are therefore usually neglected; i.e., moments of the c-field are interpreted directly as moments of the quantum field, which is of course a reasonable approximation in the classical-field limit in which the SPGPE is derived.  Moreover, individual trajectories are often interpreted as `typical' realisations of a particular experiment --- see Ref.~\cite{blakie_bradley_08} for further discussion.

Although fundamentally describing an open dynamical system, the SPGPE can be used to non-perturbatively calculate the equilibrium properties of finite-temperature Bose gases.  Once any transient dynamics have subsided, the SPGPE trajectories sample the grand-canonical equilibrium of the system in the classical-field limit, in the spirit of the well-known Langevin equation approach to sampling thermal distributions~\cite{binney_book_92}, and have been utilised, for example, to calculate the thermodynamic properties of trapped 1D Bose gases~\cite{davis_blakie_12}.  Dynamical calculations using the SPGPE method have mostly focussed on the decay of vortices~\cite{rooney_bradley_10,rooney_blakie_11} and the formation of condensates~\cite{bradley_gardiner_08,weiler_neely_08}.   

\subsection{Numerical Implementation of the SPGPE}
Applications of the SPGPE have thus far all focussed on a numerically expedient simplification of the full formalism --- the `simple growth SPGPE' --- which utilises two further approximations:  First, the growth rate $G(\mathbf{x})$ is basically constant over the bulk of the c-field region~\cite{bradley_gardiner_08}, so we neglect its spatial variation.  Second, the scattering terms appearing in~Eq.~(\ref{eq:SGPE_line3}) are neglected, on the basis that these terms are small unless large c-field currents are present, which does not occur near equilibrium.   

Numerically, the non-local nature of the scattering term introduces significant complexity, and as the associated noise is multiplicative, first- or second-order (and hence inefficient) integration algorithms would be required for stochastic convergence.   The simple growth SPGPE is, by contrast, an SDE with additive noise, which are typically more stable than SDEs with multiplicative noise~\cite{milstein_tretyakov_book_04}.  We have integrated the simple growth SPGPE using an interaction-picture method~\cite{gardiner_book_04}, and a fourth/fifth-order Runge-Kutta algorithm with an adaptive step size~\cite{press_teukolsky_book_02}.  Such an algorithm is not strictly convergent for SDEs, but in the limit of weak noise  it is an acceptable approximation \cite{milstein_tretyakov_book_04}.  

The implementation  of the SPGPE is greatly simplified if  the c-field $\Phi(\mathbf{r})$ is expanded over the single-particle basis in which the projection operator $\mathcal{P}_{\mathbf{C}}$ is defined.  In order for the cutoff defined in terms of the basis $\{\varphi_n(\mathbf{r})\}$ to be sensible, the highest-energy excitations in the $\mathbf{C}$~region should be single-particle-like (i.e., relatively unaffected by interactions).  Fortunately, for many situations of interest this condition can be satisfied simultaneously with the classical-field condition $\langle \hat{N}_k\rangle~\gg~1$~\cite{gardiner_zoller_98,gardiner_davis_03,blakie_bradley_08}.   

Many of the terms in the SPGPE are spatially local, and so efficient transformations between the modal representation and position space are required.  For homogeneous systems a plane-wave basis is appropriate, and the necessary transformations can be implemented using fast Fourier transforms~\cite{davis_morgan_02}.  For  harmonically trapped systems it is possible to use Gaussian quadrature methods to effect numerically exact transformations~\cite{blakie_davis_05a,blakie_08}.  Although the computational cost of these transformations scales (in 3D) as $M^4$ (where $M$ is the number of modes in each dimension) as opposed to the $M^3 \log M$ scaling of the fast Fourier transform~\cite{dion_cances_04,blakie_08}, the value of $M$ required is somewhat smaller than is needed to accurately represent the same harmonic oscillator states on a Cartesian grid~\cite{wright_thesis_10}.

\subsection{Projected Gross--Pitaevskii Equation (PGPE)}
The projected Gross--Pitaevskii equation (PGPE) is a simplification of the SPGPE obtained by retaining the high-occupation validity condition $\langle \hat{N}_k \rangle \gg 1$ for the modes of the c-field region, but neglecting their coupling to the $\mathbf{I}$ region.  This results in a closed system, and the resulting Hamiltonian evolution of the field $\Phi(\mathbf{r})$ conserves the energy, normalisation, and any other first integrals which may be present, such as the momentum \cite{davis_morgan_01, davis_morgan_02}, angular momentum~\cite{wright_bradley_09}, or spinor-gas magnetisation~\cite{pietila_simula_10}.

The PGPE has mostly been used for calculating the equilibrium properties of Bose gases at finite temperature~\cite{davis_morgan_01,davis_morgan_02,davis_morgan_03,blakie_davis_05a,davis_blakie_06,bezett_toth_08,bezett_blakie_09a,bisset_davis_09,wright_bradley_09,foster_blakie_10,pietila_simula_10,wright_proukakis_11}.  In practice, the PGPE seems to generate a suitable ergodic time evolution; i.e., given an arbitrary initial state, the trajectory of the field densely covers the microcanonical ensemble defined by the conserved energy and other  first integrals~\cite{sethna_book_09}. Once equilibrium is established, time averages of the field can be substituted for thermodynamic ensemble averages~\cite{davis_morgan_01,davis_morgan_02,goral_gajda_01a}.  Rugh has shown generically~\cite{rugh_97} that derivatives of the classical microcanonical entropy can be reformulated as averages of functions of the phase-space variables over the appropriate constant-energy surface.  This methodology has been applied to the PGPE, and allows for the rigorous calculation of the temperature and chemical potential of the system by time averaging~\cite{davis_morgan_03,davis_blakie_05}.  The above-cutoff region can then be described in a mean-field approximation~\cite{davis_blakie_06}. 

Because the PGPE neglects collisional processes that transfer population between the coherent and incoherent regions, it is likely to underestimate damping rates (for example), and the PGPE dynamics are potentially sensitive to the value of the cutoff~\cite{bezett_blakie_09}.  However, dynamical calculations within a pure PGPE formalism are able to provide useful insights into the dynamics of degenerate Bose-gas systems in situations where a precise identification of the method with the full field theory is impractical~\cite{wright_ballagh_08,wright_bradley_09,wright_bradley_10}.  The PGPE has also been used to establish the connection between c-field methods and more traditional theoretical methods based on U(1) symmetry breaking~\cite{wright_blakie_10,wright_proukakis_11}.    

\subsection{Truncated Wigner Projected Gross--Pitaevskii Equation (TWPGPE)}
The truncated Wigner projected Gross--Pitaevskii equation (TWPGPE)~\cite{blakie_bradley_08}  is an implementation of the truncated Wigner approximation~\cite{steel_olsen_98,sinatra_lobo_01,sinatra_lobo_02} that is computationally  identical to the PGPE, but has a distinct regime of validity.  Whereas the c-field in the SPGPE and PGPE methods constitutes an approximation to the quantum Bose field in the sense that it describes only highly occupied field modes, the TWPGPE includes a representation of the quantum fluctuations, and can thus describe weakly occupied modes as well.   However, the formal Wigner interpretation of the trajectories quickly becomes invalidated at the high temperatures at which the SPGPE and PGPE are applicable. 
 
The TWPGPE is distinguished from the truncated Wigner method described in Refs.~\cite{steel_olsen_98,sinatra_lobo_01,sinatra_lobo_02} by the implementation of a high-energy cutoff using a projection operator.  
This projection operator formalises the coarse-grained \emph{effective-field-theory} description of the Bose field~\cite{andersen_04} underlying the description of interatomic interactions by a `contact' potential (see, e.g., Ref.~\cite{morgan_00}), and provides a rigorous basis for the addition of a \emph{finite} density of quantum fluctuations to what would otherwise be a divergent local field theory~\cite{norrie_ballagh_06}.  An appropriate choice of projection operator can mitigate some of the spurious effects of quantum noise that may arise within the TWA in two and three dimensions, by effectively reducing the number of simulated modes (and hence the amount of noise) significantly~\cite{ferris_davis_08}.

\section{Validity Issues}
A well-established validity condition for the TWA at $T=0$ is that the number of particles being simulated should be somewhat larger than the number of modes in the c-field~\cite{sinatra_lobo_02}.  Intuitively, the real population of each mode should dominate the half-quantum vacuum occupation of the mode, in order for the neglect of the quantum processes represented by third-order derivatives to be valid.  This has been formalised by Polkovnikov~\cite{polkovnikov_03}, who developed a perturbation expansion for quantum dynamics around the classical-field limit, and found that the TWA is obtained as the first-order correction to the classical (GPE) dynamics.  Norrie \emph{et al.}~\cite{norrie_ballagh_06} argued that fundamentally, the \emph{local} density of real particles should be large compared to the density of vacuum fluctuations.  As the terms neglected in the TWA are only significant where the total particle density is large, the inaccuracy of the evolution in regions of low real-particle density is therefore of comparatively little consequence for the overall accuracy of the method.

Over time the error associated with the Wigner truncation grows, and at long times, the ergodic character of the PGPE causes its solutions to thermalise to a classical microcanonical equilibrium, so that a formal Wigner interpretation of the trajectories is no longer available.  Sinatra~\emph{et al.} noted~\cite{sinatra_lobo_02} that thermalisation of the initial noise population will lead to a spuriously high equilibrium temperature, and the rates of damping processes in the field may therefore be overestimated.  However, in non-equilibrium scenarios, this effect may be of little consequence, even in simulations starting from zero temperature~\cite{wright_ballagh_08,wright_thesis_10}.  In the limit of classically thermalised fields such as those described by PGPE and SPGPE equilibria, in which the formal Wigner moment correspondences are neglected, the fundamental validity condition is the mode occupation condition $\langle \hat{N}_k \rangle \gg 1$.  In practice, this is well satisfied at temperatures ranging from just above the critical temperature $T_\mathrm{c}$ down to about $0.5 T_\mathrm{c}$~\cite{blakie_davis_05b}. 

\begin{figure*}
\centerline{
\includegraphics[height=0.30\textwidth]{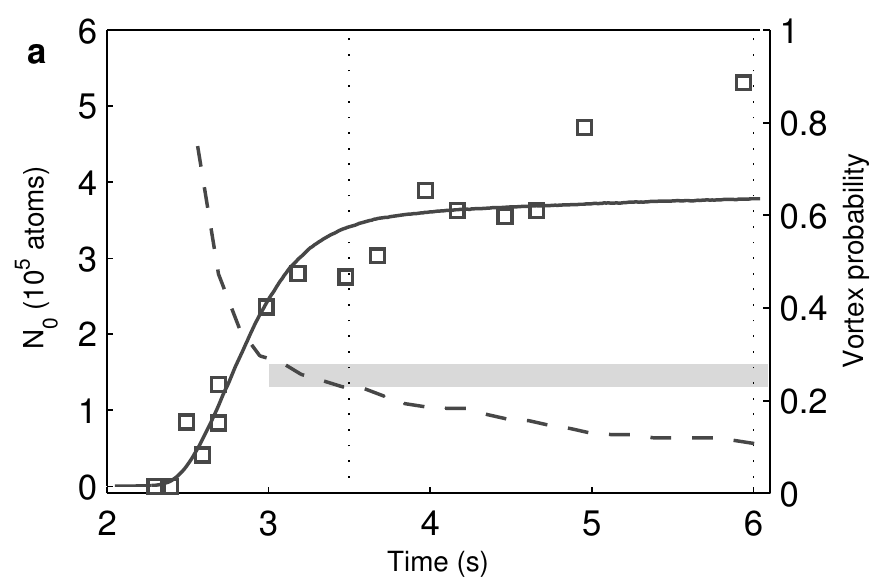}
\includegraphics[height=0.30\textwidth]{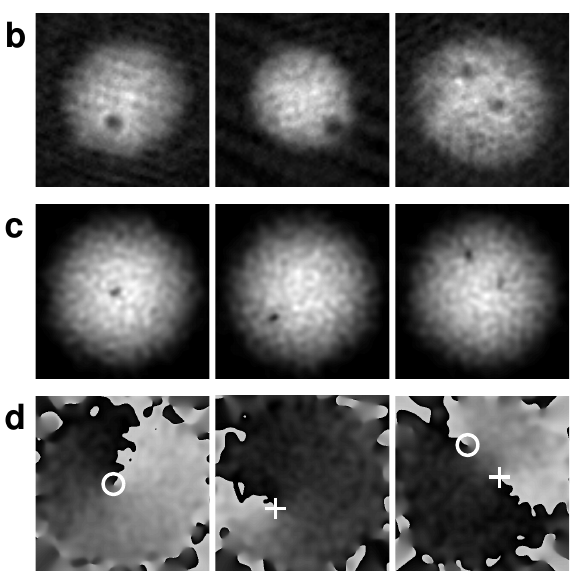}
}
\caption{Spontaneous vortices in the formation of a Bose--Einstein condensate. \textbf{a}. Squares: number of condensate atoms as a function of time measured in the experiment.  Solid line: Condensate number from SPGPE simulations with parameters: $T_\mathrm{i} =  45 $ nK, $T_\mathrm{f} = 34 $ nK, $\mu_\mathrm{i} = \hbar\omega_r$, $\mu_\mathrm{f} = 25 \hbar \omega_r$, $\hbar G(\mathbf{r})/k_\mathrm{B} T = 0.005 $.  Dashed line: probability of finding one or more vortices in the condensate as a function of time, averaged over 298 trajectories.  The shaded area indicates the statistical uncertainty in the experimentally measured vortex probability at $t= 6.0$ s.  It was observed in experiment that there was no discernible vortex decay between 3.5 s and 6.0 s  \textbf{b}. Experimental absorption images taken after 59 ms time-of-flight showing the presence of vortices.  \textbf{c}. Simulated in-trap column densities  at $t=3.5$ s (indicated by the vertical dotted line in \textbf{a}.) \textbf{d}. Phase images through the $z=0$ plane, with plusses (open circles) representing vortices with positive (negative) circulation. 
Adapted with permission from C.~N.~Weiler \emph{et al.}, {\em Spontaneous vortices in the formation of  Bose--Einstein condensates}, Nature {\bf 455}, 948 (2008)~\cite{weiler_neely_08}.} 
\label{davis_fig1}
\end{figure*}

\subsection{Relevance to Other Theories}
The SPGPE method is closely related to the stochastic GPE derived by Stoof \cite{stoof_99} within a Keldysh path-integral formalism.  In Stoof's approach the effects of dissipation, which are described using quantum-optical tools in the derivation of the SPGPE, appear as self-energy terms (cf.~Ref.~\cite{fogedby_93}).  Proukakis and Jackson \cite{proukakis_jackson_08} have shown that the Stoof SGPE is functionally equivalent to the SPGPE, when the scattering term and projection operator are neglected.  However, the precise relationship between the two formulations remains somewhat unclear, as it does not appear that a strict division between the classical modes described by the Stoof SGPE and the modes that form the `reservoir' to which the classical field are coupled is made in the Stoof approach.  Despite these differences, comparisons of numerical results of the simple growth SPGPE and the Stoof SGPE show close agreement in equilibrium~\cite{cockburn_lee_12}. 

Ergodic classical-field simulations for equilibrium correlations, similar to the PGPE, have been considered by other authors~\cite{goral_gajda_01a,schmidt_goral_03,brewczyk_gajda_07,sinatra_castin_07,sinatra_castin_08}.  These simulations typically do not include an explicit projector, leaving the cutoff to be determined by the spatial resolution of the numerical grid.  However, calculating the nonlinear term of the GPE on such a grid may lead to numerical aliasing --- a well-known issue in computational physics~\cite{gottlieb_orszag_book_77} --- particularly as the classical-field approximation \emph{requires} that all simulated modes are significantly occupied.  These authors also employ somewhat \emph{ad hoc} refinements to estimate Bose-field correlations from a classical-field model.  For example, Sinatra and co-workers~\cite{sinatra_castin_07, sinatra_castin_08} have suggested modifying the single-particle Hamiltonian in the interacting classical-field equation of motion, such that the equipartition of system energy would produce a bosonic number distribution in the ideal-gas limit.  Similarly, Brewczyk, Gajda, Rz\c{a}\.zewski and co-workers \cite{brewczyk_gajda_07} have attempted to define an `optimal momentum cutoff', by adjusting the computational grid spacing in their simulations of the interacting Bose gas, such that the occupations of the condensate or other modes match those of the ideal gas at the same temperature \cite{zawitkowski_brewczyk_04, witkowska_gajda_09}.  In our view, the division of the Bose field into classical and incoherent regions, implemented by means of a projection operator that properly restricts the classical-field approximation to modes of the field for which it is valid, and allows for a quantitative mean-field description of the complementary high-energy part of the field, is the only classical-field approach which facilitates truly quantitative descriptions of experimental systems (see, e.g., Ref.~\cite{davis_blakie_06}). 

The ZNG method for finite-temperature Bose--Einstein condensates described by Zaremba, Nikuni, and Griffin \cite{zaremba_nikuni_99} bears some superficial similarity to the SPGPE approach, in that it involves a generalised Gross-Pitaevskii equation which features damping terms representing the effects of the thermal cloud.  However, the ZNG approach is based on the fundamentally different perspective of \emph{assuming} the existence of a well-defined condensate mode, which is introduced via a symmetry-breaking ansatz~\cite{beliaev_58}.  By contrast, in the SPGPE approach no such assumption of condensation is made, and the condensate (if one is present) must be extracted \emph{a posteriori} from the field correlations \cite{goral_gajda_02, blakie_davis_05a}.  Further discussion of the different philosophies of the c-field and ZNG approaches can be found in Refs.~\cite{griffin_zaremba_chapter_12} and~\cite{wright_davis_chapter_12}.  The main advantage that the ZNG method offers is a description of the coupled dynamics of the condensate and the full thermal cloud, in which the kinetics of the latter are modelled in a Boltzmann equation approach.  Such a coupled condensate-cloud description appears to be essential for accurately describing certain collective oscillations of the gas at high temperatures \cite{jackson_zaremba_02c, morgan_rusch_03, bezett_blakie_09}.  However, as it is based on the assumption that a well-defined condensate exists, the ZNG approach is not applicable to more general scenarios involving low-dimensional systems \cite{bisset_davis_09,davis_blakie_12,alkhawaja_andersen_02a,cockburn_gallucci_11}, regimes of turbulent matter-wave dynamics \cite{lobo_sinatra_04,wright_ballagh_08,berloff_svistunov_02}, or non-equilibrium passage through the transition to condensation \cite{weiler_neely_08, bradley_gardiner_08,stoof_bijlsma_01}.

\section{Application}
The first experiments on 3D condensate formation observed BECs to grow essentially quasistatically as smooth Gross-Pitaevskii-like states~\cite{miesner_stamper-kurn_98,kohl_davis_02}, and were successfully modelled using kinetic-theory approaches~\cite{davis_gardiner_00,bijlsma_zaremba_00,davis_gardiner_02}.  Later experiments by Shvarchuck \emph{et al.}~\cite{shvarchuck_buggle_02} and Hugbart \emph{et al.}~\cite{hugbart_retter_07} studied the formation dynamics of non-equilibrium quasicondensates.  More recently, the Anderson group at the University of Arizona evaporatively cooled Bose gases rapidly from near degeneracy to BEC in a weakly oblate harmonic trap~\cite{weiler_neely_08}.  When imaging along the vertical ($z$) direction, they found that approximately 25\% of BEC column-density images featured holes consistent with vortices in the condensate.  

Simulating the Anderson group's experiments with the simple growth SPGPE has helped elucidate the phenomenon of spontaneous vortex formation~\cite{weiler_neely_08}.  The initial states for the simulations were sampled from an equilibrium ensemble near degeneracy, with a temperature $T$ and chemical potential $\mu$.  Evaporative cooling was modelled by  suddenly changing the bath parameters $T$ and $\mu$ to experimental values measured immediately after the most rapid phase of condensate growth.  The dimensionless growth rate $\hbar G(\mathbf{r})/k_\mathrm{B} T$ was set to a constant in time and space chosen so that the simulated condensate growth curve matched that observed experimentally, as shown in Fig.~\ref{davis_fig1}.  A total of 298 trajectories were simulated, with each interpreted as a separate experimental run.

The simulated probability of observing a single vortex in the resulting condensate agrees well with the experimental results, as indicated in Fig.~\ref{davis_fig1}.  The majority of spontaneously formed vortices aligned with the $z$~axis, and hence were easily observed in the column-density images.  The vortices were found to live for many seconds in the experiment, somewhat longer than  in the simulations.  However, the temperatures ultimately reached in the experiment are below those for which the SPGPE is applicable, and so this additional cooling after the initial growth phase was not simulated.  The discrepancy between experimentally measured vortex lifetimes and the SPGPE simulation results is therefore consistent with the dependence of vortex decay rates on the system temperature~\cite{fedichev_shlyapnikov_99,duine_leurs_04,jackson_proukakis_09,rooney_bradley_10}.  These results demonstrate that the SPGPE is a powerful tool for the quantitative modelling of the non-equilibrium dynamics of Bose gases at finite temperature.

\section{Relevance to Other Systems}
The SPGPE could potentially be useful for simulating non-equilibrium exciton-polariton systems, which are only ever in steady state due to the pumping and finite lifetime of the quasiparticles (see, e.g., Ref.~\cite{deng_haug_10}).  However, a more general representation of the $\mathbf{I}$ region would likely be necessary, as the reservoir will not in general be in equilibrium, and the scattering term neglected in the simple growth SPGPE could play an important role in establishing the steady state of the $\mathbf{C}$ region.  Wouters and Savona have applied methodology similar to the SPGPE~\cite{wouters_savona_09} to investigate superfluidity in polariton systems~\cite{wouters_savona_10}.

\section*{Acknowledgements}
\vspace{-\baselineskip}
The authors acknowledge the support of the Australian Research Council Discovery Projects DP1094025 and DP110101047,  FRST contracts  NERF-UOOX0703 and UOOX0801, Marsden Fund contracts UOO509, UOO0924, and UOO10106, and the Royal Society of New Zealand RDF-UOO1002.  

\bibliographystyle{prsty}

\end{document}